\documentclass[
final            
  ,numberedheadings 
,sort&compress]{aipproc}

\layoutstyle{6x9}

\newcommand{\beq}{\begin{equation}}
\newcommand{\eeq}{\end{equation}}
\newcommand{\be}[1]{\begin{equation}\label{#1}}
\newcommand{\ee}{\end{equation}}

\newcommand{\bea}{\begin{eqnarray}}

\newcommand{\eea}{\end{eqnarray}}

\newcommand{\bm}[1]{\mbox{\boldmath{$ #1$}}}
\newcommand{\Tr}{\mbox{Tr}\,}

\renewcommand{\det}{\mbox{\rm det}}
\newcommand{\Det}{\mbox{\rm Det}\, }

\newcommand{\MatrixII}[4]{
   \pmatrix{ {#1}  &  {#2} \cr
             {#3}  &  {#4} \cr} }

\newcommand{\refeq}  [1] {(\ref{#1})}

\newcommand{\reffig} [1] {fig.~\ref{#1}}

\newcommand{\refsect}[1] {sect.~\ref{#1}}

\newcommand{\cluster} {{\sf M}}
\newcommand{\kernel} {{\sf A}}

\newcommand{\singleScat}[2] {\bigl[ {\sf S}^{(1)#1}_{#2} \bigr]}
\newcommand{\multiScat} {{\sf S}}
\newcommand{\ToInterior}{{\sf C}}
\newcommand{\ToInfinity}{{\sf D}}

\newcommand{\scatcoefOrb}{{\alpha}}

\newcommand{\stab} {\mathbf{J}}

\newcommand{\identityMat} {{\sf {1}  }}

\newcommand{\tracMat}[3] { \bigl[ {\sf t}_{#1}^{(#2) #3} \bigr]}

\newcommand{\fredholm} {{ F}}

\begin{document}

\title{Periodic orbits in scattering 
from elastic voids}


\author{Niels S\o ndergaard}{
  address={Division of Mathematical Physics, LTH, Lunds Universitet, Sweden}
}

\author{Predrag Cvitanovi\'{c}}{
  address={Center for Nonlinear Science, School of Physics, Georgia Institute of Technology, Atlanta, USA}
}

\author{Andreas Wirzba}{
  address={Institut f\"ur Kernphysik (Theorie),
       Forschungszentrum J\"ulich,  J\"ulich, Germany} 
}

\begin{abstract}
  The scattering determinant for the scattering of waves from several
  obstacles is considered in the case of elastic solids with voids. The
  multi-scattering determinant 
   displays contributions from periodic ray-splitting
  orbits. A discussion of the weights of such orbits is presented.
\end{abstract}

\maketitle


\section{Introduction}
Eventhough the word {\em scattering}  appears to imply 
{\em transport}, 
studies of Helmholtz
scatterers have shown effects of {\em trapped} periodic orbits
\cite{Predrag}. As there is nothing particular about the Helmholtz
equation, 
similar effects are expected for wave equations for other media such
as dielectric or elastodynamic ones.

We shall discuss the relation between periodic trapped rays and  
the scattering determinant corresponding to a medium with several 
polarizations, each with their
own velocity. The example to be treated is 
the case of elastic wave propagation
in a solid punctured by  a finite number of voids.  These systems have
the feature that
a ray hitting a
boundary can either reflect or refract. Particularly, ray splitting occurs 
when the polarization changes. This leads to a ray dynamics which no longer is
unique, since -- in general --  
a single polarized ray evolves into a tree of rays. A
similar behaviour is observed in microwave resonators with dielectrica: 
characteristically
rays can either be reflected or be transmitted 
at the boundaries of the dielectrica.

\section{Scalar case}
\label{sect:ScalarCase}
The Helmholtz equation 
\beq 
  (\Delta + k^2) \psi = 0 
\eeq 
describes the wave
propagation of a scalar field $\psi$ of wave number $k$ 
in a homogeneous and isotropic 
medium. Furthermore, if 
obstacles are embedded in this background,
the scalar field typically has to satisfy
Dirichlet ($ \psi = 0$) or Neumann ($\partial \psi/\partial n = 0$) 
boundary conditions on the surfaces of the obstacles.  
For an exterior problem, i.e.\ a scattering problem, 
the geometry of the scatterers
and the corresponding boundary conditions are usually 
specified and the typical goal is
to calculate the scattering matrix $\multiScat$. In short, this matrix
contains the information on how an incoming  wave transforms to
a superposition of outgoing scattering solutions. 
From the knowledge of the scattering matrix several
interesting quantities can be calculated: cross-sections, resonances, 
phase shifts, time-delays etc.

One way of finding the scattering matrix is via the so-called {\em null-field
  method}
\cite{lloyd,lloyd_smith,berry81,gaspard,AW_report,PetersonStrom,bostrom,hwg97,cavityThesis}.
A given field on the  boundary  of one scatterer 
gives rise to secondary fields 
on the full set of
boundaries, including the one at infinity. These fields are calculated via
boundary integral identities. If a basis is chosen for each boundary, the
initial and the secondary fields can be expanded in these bases.
The matrices that describe their relationships can be used to 
construct the
scattering matrix~\cite{gaspard}. In this way
the application of the standard null-field method 
determines the $\multiScat$ matrix as
\beq
\label{scatCMD}
\multiScat = \identityMat - i \, \ToInterior \, \cluster^{-1} \ToInfinity, 
\eeq
where the $\ToInterior$ and $\ToInfinity$ matrices connect the incoming,
respectively, outgoing waves 
to the interior scattering boundaries and where the matrix
$\cluster$ relates the waves at one interior scattering boundary to
the waves at a different one.
$\cluster$ itself can be written in terms of a transfer matrix
$\kernel$ 
as
\beq
\cluster=\identityMat + \kernel .
\eeq
The multi-scattering expansion arises when 
\beq
\cluster^{-1} = \identityMat - \kernel + \kernel^2 + \dots
\eeq
is inserted in \refeq{scatCMD} 
\cite{lloyd,lloyd_smith,berry81,gaspard}.
This signals that $\cluster$ has the role of
an inverse multi-scattering matrix.
In some situations the exact matrix elements of
these infinite dimensional matrices, $\kernel$, $\ToInterior$, and 
$\ToInfinity$,  can be derived. Of course 
these matrices
depend crucially on the scattering problem in question, but the overall
structure is the same. In general, the null--field method can be applied to
solve  numerous multi-scattering problems in various media.

An explicit example is the scattering of a scalar wave, 
propagating in the two-dimensional plane, 
off  a finite number of non-overlapping 
hard discs that have fixed positions and Dirichlet boundary conditions. 
For the discussion here only the matrix
$\cluster$ is of interest. The fields at two 
different discs $j$ and $j'$, expanded in
basis states $\exp(i m \phi_j)$ and $\exp(i m' \phi_{j'})$ (in polar
coordinates with $m,m'=0,\pm 1,\pm 2,\cdots$  two-dimensional 
angular momentum quantum numbers) are connected via the (transfer) 
kernel~\cite{gaspard,AW_report,aw_chaos,aw_nucl,wh98,threeInARow}  
\beq 
\label{scalarCluster}
[\kernel]_{m m'}^{j j'} = (1-\delta_{j j'}) \frac{J_m (k
a_j)}{H_{m'}^{(1)}(k a_{j'})} \, H_{m-m'}^{(1)}(k R_{j j'}) \, e^{i m
\alpha_{j'}^{(j)}- i m'(\alpha_{j}^{(j')}-\pi)} \,.  \eeq Here
$R_{jj'}$ is the center-to-center distance of the discs of radii $a_j$,
and $\alpha_{j'}^{(j)}$ is the angle to the center of cavity $j'$ in the
coordinate system of cavity $j$. All this follows from 
the application of the above-sketched 
null--field method to the scalar-wave problem. 
Less familiar is the ray limit of the pertinent
multi-scattering determinant $\Det \cluster(k)$ 
\cite{gaspard,AW_report}.  As the various
building blocks of the scattering matrix are known in analytical form
it is possible to explicitly calculate the short-wave-length
limit. The main result
is that asymptotically~\cite{AW_report} 
\beq
\label{ScalarFredGeom}
\Det \cluster(k) \approx F =\left. \exp \left(- \sum_p \sum_{r = 1}^\infty
\frac{1}{r} \, \frac{e^{i r \,( k L_p-\nu_p\pi/2)
}}{|\Det(\identityMat-\stab_p^r)|^{1/2}}\, z^{r n_p} \right)\right|_{z=1}  , 
\eeq 
where
$p$ is a prime orbit (a periodic orbit that cannot be split into 
shorter ones), $r$ is the number of its repeats,
$L_p$ is its length, $\nu_p$ is its Maslov index, 
$\stab_p$ is its monodromy matrix,
and $n_p$  is its number of bounces on the
scatterers. The right-hand side of \refeq{ScalarFredGeom} is the
semiclassical {\em spectral} determinant~\cite{Predrag,gutbook}.
The parameter $z$, which at the end is  put equal to unity, serves
in the formal  expansion of $F$ in $z$ up to a sufficiently high 
number $N_{max} \geq
n_p$. 
A {\em periodic orbit} is defined here as a 
closed ray in phase space obeying
the law of reflection at the scatterers. Our goal is to generalize 
these types of problems to two-dimensional elastodynamics.

\section{Elastodynamics}
In isotropic elasticity the wave equation in the frequency domain $\omega$
is
\beq
\mu \Delta\, {\bf u} + (\lambda + \mu) {\bf \nabla ( \nabla \cdot u)} +
\rho \omega^2 {\bf u} = 0 \,,
\label{pde}
\eeq where ${\bf u}({\bf x})$ is the displacement vector field in the body,
$\lambda$, $\mu$ are the material-dependent Lam\'e coefficients and
$\rho$ is the density \cite{lAndl,auld}. This wave equation admits two
different polarizations: longitudinal L  and transverse T with velocities
\beq 
c_L = \sqrt{\frac{\lambda + 2 \mu}{\rho}} \qquad \mathrm{and}
\qquad c_T = \sqrt{\frac{\mu}{\rho}} \,.  
\eeq
The longitudinal and transverse waves correspond to {\em pressure}, 
respectively, 
{\em shear} deformations 
(or to {\em primary} and {\em secondary} arriving pulses in seismology). This
leads to the law of refraction for incoming plane waves
 \be{Snell}
\frac{c_L}{c_T}=\frac{\sin \theta_L}{\sin \theta_T},
\ee
where $\theta_L$, $\theta_T$ denote the angle of incidence or reflection
of the longitudinal and transverse wave, respectively, 
measured with respect to the
normal to the surface.

The stress tensor in elasticity has the form
\beq
\sigma_{ij} = \lambda \, \partial_k u_k \delta_{ij} +
\mu  \left(\partial_i u_j + \partial_j u_i\right) \,.
\eeq
The  boundary conditions considered here 
are {\it free}. Hence
\be{freebound}
{\bf t}({\bf u}) \equiv \bm{\sigma}({\bf u}) \cdot {\bf n} 
=
 \biggl[\lambda \Bigl({\bf \nabla} \cdot \bf{u}\Bigr)
{\bf 1}
 + \mu \Bigl\{ \bigl({\bf \nabla} \bf{u}\bigr)
 + \bigl({\bf \nabla}
 \bf{u}\bigr)^{\mathsf{T}}\Bigr\}\biggr]
 \cdot {\bf{n}} = 
{\bf 0}
\ee
for the displacement field at the boundary 
where $\mathbf{ n}$ denotes the normal to
the boundary and $\mathsf{T}$ indicates a transposition. 
The operator $\mathbf{ t}$ refers to the traction.

 \section{Scattering determinant}
\begin{figure}[t]
 \rotatebox{-90}{ \includegraphics[width=.5\textwidth]{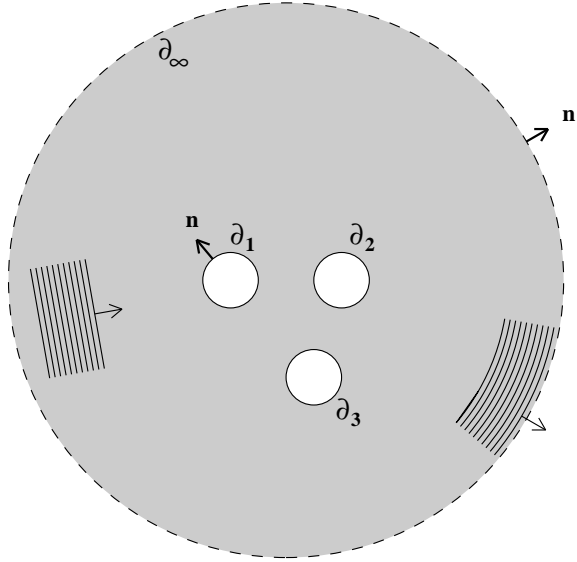}}
  \caption{\label{scatZonePl} General zone of two--dimensional cavity
    scattering. 
   Here the unit vectors ${\bf n}$ denote  the outside normals
       to the surfaces
       $\partial_i$ ($i=1,2,3$) of the cavities or to  the surface
       $\partial_\infty$ of the total scattering zone. Furthermore,
       an incoming plane wave and a scattered cylindrical wave are shown.}
\end{figure}

As mentioned, in this treatment
the medium corresponds to a cylindrical solid made, for simplicity, of
an isotropic and homogeneous elastic material.  The scattering
geometry consists of parallel cylindrical voids, which are
perpendicular to the endcaps of the overall cylinder, see
\reffig{scatZonePl}.  If the fields are stimulated by in-phase 
line-sources parallel to the voids, this symmetry is
respected and the problem reduces to one of two-dimensional
elasticity, see Fig. 1, referred to as {\em plane strain}. This 
scattering problem generalizes 
the simpler scalar scattering off discs in two dimensions, 
mentioned in \refsect{sect:ScalarCase}.

Similar to the scalar case \cite{AW_report,wh98,hwg97}, the scattering
determinant may be factorized into an incoherent single-scatterer part
(in terms of the determinants over the 
single-scattering matrices  ${\sf S}^{(1)j}(\omega)$, 
$j=1,2,\cdots$) and
the genuine multi-scatterer part (in terms of the determinant of the 
inverse multi-scattering
matrix $\cluster(\omega)$) \cite{cavityThesis}
\begin{equation}
\label{detFactor}
 {\det \,} {\multiScat}(\omega)
 = \left\{\! \prod_{{j} 
 \in \mathrm{Cavities}}\det{} {\bigl[ {\sf S}^{(1)j}(\omega) \bigr]}  \!\right \}
  \frac{ {\Det}\bigl[{\cluster}(\omega^\ast)^\dagger \bigr ]}
 { {\Det}\bigl[{\cluster}(\omega) \bigr]}.
\end{equation}
When the {\em relative} positions of the cavities are changed, only the
latter factor, composed from the {\em cluster} determinant $\Det \,
      {\sf M}(\omega)$, changes: 
\beq
\label{clusterDef}
{\sf M} = {\bf 1}+ {\sf A}\,,
\eeq
\begin{eqnarray*}
 &&\Bigl[{\kernel}^{jj'}_{ll'}\Bigr]_{i i'} = (1\!-\!\delta_{jj'})
  \frac{a_j}{a_{j'}} \sum_{\sigma=L}^T\sum_{\sigma'=L}^T
\bigl[ {\sf t}_l^{(J) j} \bigr]_{i \sigma}
  \Bigl[{\sf T}^{(+) jj'}_{ll'}\Bigr]_{\sigma\sigma'}
  \bigl[ {\sf t}_{l'}^{(+) j'} \bigr]^{-1}_{\sigma' i'}\,,
 \\
 &&\Bigl[ {\sf T}^{(+) jj'}_{ll'}\Bigr]_{\sigma \sigma'}=
  \delta_{\sigma \sigma'} 
  H_{l-l'}^{(+)}(k_{\sigma} R_{jj'}) e^{il \alpha_{j'}^{(j)}-
  il'(\alpha_{j}^{(j')}-\pi)} .
\end{eqnarray*}
The indices $i,i' \in \{ r,\phi \}$ are the labels of 
the two-dimensional  polar coordinates
and the indices $\sigma,\sigma' \in \{ L,T \}$ refer to  
the two polarization states. 
The diagonal 
matrix $\Bigl[ {\sf T}^{(+) jj'}_{ll'}\Bigr]$ may be
interpreted as a translation matrix acting on the polarized 
scattering states, where $k_L=\omega/c_L$ and $k_T=\omega/c_T$
are the 
pertinent wave numbers~\cite{PetersonStrom,bostrom}.
As in \refeq{scalarCluster}, $R_{jj'}$ is
the center-to-center distance of the circular cavities of radii $a_j$
and $\alpha_{j'}^{(j)}$ is the angle to the center of cavity $j'$ in the
coordinate system of cavity $j$. 
The single-cavity scattering matrices (enumerated by the
cavity index $j=1,2,\dots$) are separable in the
(two-dimensional) angular momentum $l=0,\pm 1,\pm 2, \cdots$ 
due to
the rotational symmetry. They have the general form: 
\beq
\label{singleCavityDef}
\singleScat{j}{l} = -\tracMat{l}{+}{j}^{-1} \cdot \tracMat{l}{-}{j}
\eeq
in terms of the $2\times 2$ traction {\em matrices} $\tracMat{l}{Z}{j}$ which
incorporate the {\em free} boundary conditions \refeq{freebound}.
 Here  the ``type'' $Z \in \{+,-,J \}$ 
refers to outgoing, incoming or regular scattering states and involves
$H_l^{(1)}$, $H_l^{(2)}$ or $J_l$ Bessel functions of argument
$z_{L,T}\equiv a k_{L,T}$, 
respectively. Thus we have, e.g., for the outgoing case:
\beq
\label{tracOut} \small
{\tracMat{l}{+}{j}}=\frac{2 \mu}{a_j^2}\,\MatrixII{(l^2
  -z_T^2/2)H^{(1)}_l(z_L)-z_L 
 \frac{d H^{(1)}_l(z_L)}{dz_L}}{i\, l \left(H^{(1)}_l(z_T) -z_T 
 \frac{d H^{(1)}_l(z_T)}{dz_T}\right)}{i \,l \left(H^{(1)}_l(z_L) -z_L 
 \frac{d H^{(1)}_l(z_L)}{dz_L}\right)}{-(l^2 -z_T^2/2) H^{(1)}_l(z_T)+z_T 
\frac{d H^{(1)}_l(z_T)}{dz_T}} \,.
\eeq
Note that
the single-cavity scattering matrix connects different polarizations. For a
full discussion, see \cite{cavityThesis,izbicki,paoAndmow,cavityLetter}. The
connection to the interior problem of a single disc is described in
\cite{disc}.
\begin{figure}[t]
  \includegraphics[width=.85\textwidth]{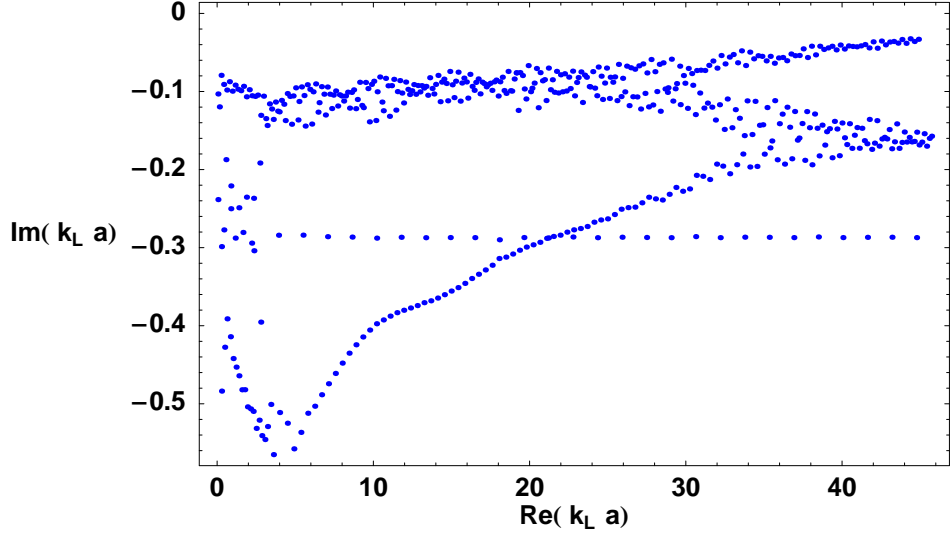}
  \caption{\label{resoPl} Elastodynamic scattering resonances 
from \cite{cavityLetter} for two cavities of common radius $a$ 
($A_1$-representation, 
center-to-center-separation $R=6a$) in the
complex  
longitudinal Helmholtz number--plane:  $k_L a =  \omega a /c_L  $. }
\end{figure}

As first shown for the scalar problem~\cite{AW_report}, 
the poles of the cluster
determinant $\Det \cluster(\omega)$ cancel -- by construction -- the poles
of the single-scattering determinants. Likewise, the poles of $\Det
\cluster(\omega^*)$ are canceled by the zeros 
of the single-scattering
determinants. Thus all scattering resonances defined by the 
{\em poles} of
the scattering determinant ${\det \,} {\multiScat}(\omega)$
can be found from the {\em zeros} of the cluster
determinant  $\Det \cluster(\omega)$
\cite{AW_report,wh98}. As an example consider the
resonances in \reffig{resoPl}, see \cite{cavityLetter}, of a
two--cavity system made of polyethylene \cite{izbicki}, 
a material with
$c_L = 1950$ m/s and $c_T = 540$ m/s. Furthermore, it is assumed that
the  cavity radii are  equal, i.e.\ $a_1=a_2\equiv a$,
and that the
inter-cavity separation $R$, measured from the centers, 
is 6 times larger than $a$. Note that
the regular spaced horizontal set of resonances in \reffig{resoPl}
is placed below an irregular set. This is opposite to the scalar
Helmholtz case for the same geometry where the regular spaced
resonances are above the irregular ones
\cite{vwr94,vattay94,threeInARow,rvw96}. The regular resonances
particular to the fundamental $A_1$-representation are well described
by the following condition \cite{aw_chaos} \beq 0 = 1 + \exp(i k_L
L)/\sqrt{\Lambda} \eeq with the length $L = 4 a$ and instability
$\Lambda =5 + 2 \sqrt{6} $ \cite{Predrag}. $L$ corresponds to the length of
the shortest periodic orbit moving in a symmetry--reduced domain 
spanned by the
surface of the cavity and the center-of-mass of the two
cavities. $\Lambda$ is obtained from the product of ray matrices as
the leading eigenvalue of the monodromy 
matrix \cite{gutbook,Predrag} of the
corresponding (geometric acoustic) ray system. 
This next raises the
question about the effect of the remaining set of orbits.

\section{Orbits in time--delay}
\begin{figure}
  \includegraphics[width =0.9\textwidth ]{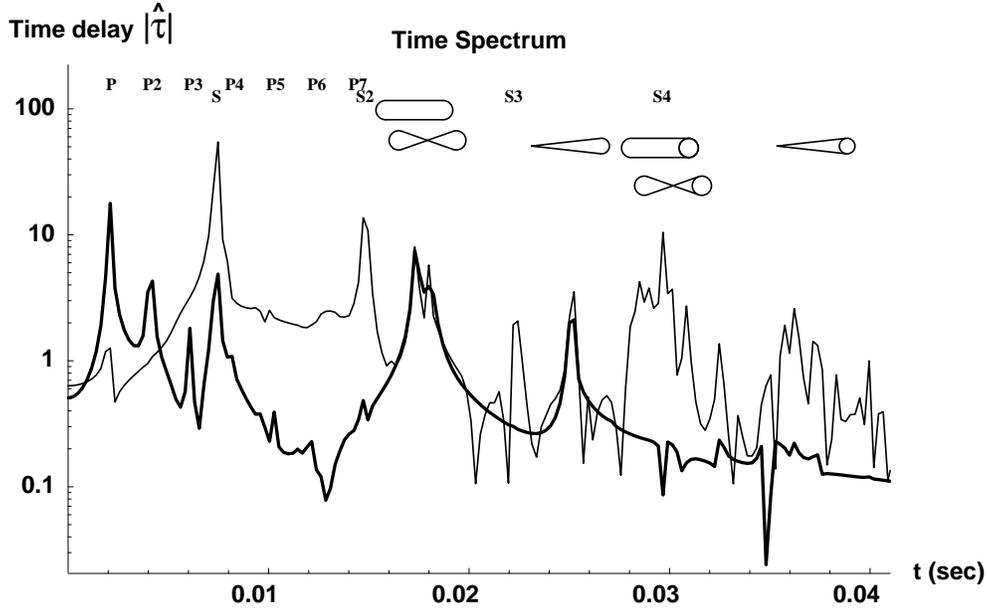}
  \caption{\label{timeSpec} Power spectrum of the time--delays derived 
from the cluster phase shift of two cavities of common radius $a$
and center-to-center separation $R=6a$. The thick and 
thin line
denote the $A_1$- and $B_2$-representation, respectively.  
The symbols P$i$ and S$i$ 
label, in turn, periodic orbits with $i$ legs  of pressure  or shear
polarization. 
The  circular arcs indicate Rayleigh--surface waves. }
\end{figure}

For real frequencies the total scattering phase $\Theta$ is given by 
the sum
over the cluster phase $\Theta_{c}$ and the single cavity phases $\Theta_j$:
\beq
\Theta(\omega) = \frac{1}{2 i} \ln \, \det \, S(\omega) \qquad \mathrm{and}
\qquad  
\Theta_c(\omega)=\frac{1}{2 i} \ln
         \frac{\Det \, M(\omega^\ast)^\dagger}
              {\Det \,  M(\omega) } \, ,
\eeq
see
\refeq{detFactor}.
Likewise, the derivative with respect to frequency $d\Theta/d\omega$, the
Wigner-Smith time delay, can be decomposed into single-scatterer and
cluster contributions. The numerics of the cluster time--delay
$d\Theta_c/d\omega$ show fluctuations which are related to trapped 
periodic orbits in
the scattering geometry, see \reffig{timeSpec} where the results for two
identical cavities (the same system as in \reffig{resoPl}) 
are presented. Due to
the symmetry of the system 
the cluster delay decomposes further into a sum over four
irreducible representations of the symmetry group $C_{2v}$ \cite{hamermesh},
of which the representations 
$A_1$ and $B_2$ are shown.  Some of these orbits are diffractive (see
\cite{cavityThesis,cavityLetter} for more details) including segments of
surface propagation of Rayleigh type, which are also important in 
earthquakes \cite{viktorov}. For
these proceedings we focus the discussion to   
purely non-diffractive contributions, called
geometrical ray-splitting orbits.

\section{Expanding the cluster determinant}
As in the scalar case \cite{AW_report}, a central point in the 
orbit-construction is 
the definition of the cluster determinant in terms of traces:
\beq
\label{eq:fredExp}
\fredholm(z)=\Det (\identityMat + z\,\kernel) \equiv \exp \left(-
\sum_{n=1}^\infty \frac{z^n \, \Tr (-\kernel)^n}{n} \right) \,, 
\eeq
where $z$ is again a formal expansion parameter which is put equal to one
at the end. Equation \refeq{eq:fredExp}  holds if $\kernel$ is
trace--class, and its Taylor expansion  is called the cumulant expansion
\cite{aw_chaos,aw_nucl, AW_report,wh98}. Trace-class operators (or
matrices) are those, in general, non-Hermitian operators (matrices) of
a separable Hilbert space which have an absolutely convergent trace in
{\em every} orthonormal basis \cite{reed_simon,simon}. Especially the
determinant $\Det (\identityMat + z \kernel)$ exists and is an entire
function of $z$, if $\kernel$ is trace-class.  Presently the trace-class
property of $\kernel$ has only been proved in detail in the
two-dimensional scalar case \cite{AW_report,wh98} and sketched for the
three-dimensional case in \cite{hwg97}. Nevertheless, we shall proceed
as if this is true also in our elastodynamical case.  This is supported
by the following numerical evidences:
(i) the sum over the moduli of the eigenvalues of
the matrix $\kernel$ from \refeq{clusterDef} is absolutely converging in
agreement with the expected trace-class property of this matrix; (ii)
the determinant \refeq{eq:fredExp} converges to a finite result as the
dimension of $\kernel$ is increased beyond a minimal number
$N_{\rm dim} > 2\times \left(\frac{e}{2} (c_L/c_T) |k_L| a \right)$, see
\cite{cavityLetter} and also \cite{AW_report,berry81}; and (iii) the
resonances of the cumulant expansion truncated at 
fourth order $z^4$ agree very well
with the exact ones plotted in \reffig{resoPl}, see \cite{cavityLetter}.

\section{Ray limit and orbits}
The expansion \refeq{eq:fredExp}
indicates that the cluster determinant can be obtained from the
knowledge of an increasing number of traces.  Moreover, each trace
$\Tr \kernel^n$ is given by the sum over all exact periodic {\em
itineraries} of topological length $n$, see \cite{AW_report}.  In the
saddle-point approximation the periodic itineraries become the
periodic orbits of topological length $n$, which means that they
bounce $n$ times between the cavities \cite{AW_report}.  These orbits
fulfill the laws of reflection and refraction and have phases
corresponding to their time periods of revolution $T_p$.

A periodic itinerary of topological length $n$ corresponds to
a  cyclic product of $n$ terms   
involving  one operator of the type
\beq
 \tracMat{l}{+}{j}^{-1} \cdot \tracMat{l}{J}{j}
\eeq
(which corresponds to the ${\sf T}$-matrix-part (times $-1/2$) 
of a
single-cavity scattering matrix
\refeq{singleCavityDef}) followed by one
translation operator $\Bigl[ {\sf T}^{(+) jj'}_{ll'}\Bigr]$ 
(the free propagation). 
The pruning rule that two
successive scatterings must take place at different cavities 
is automatically built in (see
the term $(1-\delta_{jj'})$ in the kernel $\kernel$). 
The ray limit of
the single cavity ${\sf T}$-matrix  
gives unitary reflection
coefficients similar to those of the scattering from an infinite
half--plane \cite{lAndl,disc}. 
This leads to an overall amplitude $\alpha_p$ defined 
as a product over all reflection coefficients along the orbit. This amplitude
describes the leakage from the orbit due to ray splitting.

The calculation of the geometric amplitudes of the orbits requires
more work.  See \cite{shudo} for a general discussion with respect to
the interior scalar case and \cite{AW_report} for the exterior counter
part. Asymptotic wave theory indicates
\cite{kellerPTD,KellerElasto,rulf,achenbach} that for {\it open}
trajectories in two dimensions the amplitude scales as $(k R)^{-1/2}$
where $k$ is the wave number in question and $R$ is the radius of
curvature of the wave front at the observer. This radius is studied in
e.g.\ geometric optics. It it is possible to keep track of its
evolution in the free-propagation period between the scatterers and at
impacts including possible refractions with the help of suitable ray
matrices \cite{cavityThesis}.  Indeed, for our problem it can be shown 
that all those open segments that have fixed end points, but
intermediate points (variables) determined by saddle-point
integrations, have such an amplitude evolution. This comes about by
calculating the accompanying sparse Hessian of this restricted
integration.

For a full saddle-point integration over all variables, in other
words for a {\it periodic}
orbit $p$, the amplitude turns out to be expressible as yet another sparse
Hessian that can be expanded into Hessians of the type of the
previously considered open pieces; see \cite{AW_report} for
the scalar case. The use of the previous information then allows 
the full calculation with 
the  amplitude evolving as
\beq 
\mathcal{A}_p = \frac{\alpha_p}{|\Det
({\identityMat - \stab_p})|^{1/2}} \,\, z^{n_p}\,,
\eeq 
where $\stab_p$ is the product of the ray matrices and 
and $\alpha_p$ is the product of the reflection
coefficients,  calculated along the orbit $p$ with its $n_p$ 
bounces. This
form is precisely part of the conventional semiclassical density of
states \cite{gutbook,brack,Stock}. However, the formal parameter $z$
is also present and can be seen as a counting and ordering parameter
of the various orbits in the expansion over infinitely many orbits
\cite{Predrag,artuso1,artuso2}.

Incorporating the results of the geometric ray-splitting orbits gives
the following factor of the ray-dynamical approximation of the cluster
determinant: 
\beq 
\label{fredGeom} 
\fredholm_G(z) = \exp \left(-
\sum_p \sum_{r = 1}^\infty \frac{1}{r} \, \scatcoefOrb_p^r \,
\frac{e^{i r \, \omega T_p }}{|\Det(\identityMat-\stab_p^r)|^{1/2}}\, z^{r
n_p} \right) \, .  
\eeq 
The sum over $r$ counts the repeats
of the primary periodic orbits, the {\it prime} cycles $p$. If the
logarithmic derivative with respect to $\omega$ is taken, a result very
similar to the spectral density for the interior problem is
obtained. This is in agreement with the general result for the
density of states for ray-splitting systems described in
\cite{couch}. Similar results for the case of flexural vibrations in
the interior case are given in \cite{HB}.  As the orbits are unstable
and $\stab_p$ is symplectic, it is possible to expand, for each orbit,
the instability
denominator in \refeq{fredGeom} in the inverse of its leading
eigenvalue $1/\Lambda_p$ and to obtain a so-called
Gutzwiller--Voros resummed zeta function similar to those of
two-dimensional Hamiltonian flows~\cite{Predrag}: 
\beq 
 \fredholm_G(z)
 =\prod_{k=0}^\infty \, \zeta_k^{-1}(z) \, , 
\eeq 
where 
\beq 
 1/\zeta_k(z)
 = \prod_p (1-t_p^{(k)} ) \quad \mbox{with} \quad t_p^{(k)} =
 \scatcoefOrb_p \, \frac{e^{i \omega T_p}}{\sqrt{
 |\Lambda_p|}\Lambda_p^k} \, z^{n_p} \, .  
\eeq

\section{Summary}
Detailed studies of Helmholtz scattering determinants at small wave lengths
have shown the influence of periodic orbits.  The case of scattering from
voids in two--dimensional elastodynamics was considered here with a discussion
of the analytical contribution of periodic ray-splitting orbits to the
scattering determinant.


\begin{theacknowledgments}
 N.S. acknowledges discussions with J.~D.~Achenbach 
and funding from  the European Network on {\it Mathematical aspects of Quantum
  Chaos},  the Crafoord Foundation and  the Swedish Research 
Council.

\end{theacknowledgments}



\bibliographystyle{aipproc}   


\IfFileExists{\jobname.bbl}{}
 {\typeout{}
  \typeout{******************************************}
  \typeout{** Please run "bibtex \jobname" to optain}
  \typeout{** the bibliography and then re-run LaTeX}
  \typeout{** twice to fix the references!}
  \typeout{******************************************}
  \typeout{}
 }

\end{document}